\documentclass{llncs}
\usepackage[override]{cmtt}
\usepackage{graphicx}
\usepackage{macros}
\newcommand{\ket}[1]{\left| #1 \right\rangle} 
\newcommand{\bra}[1]{\left\langle #1 \right|} 
\newcommand{\oomit}[1]{}
\usepackage{listings}
\usepackage{url}

\lstdefinelanguage{isa} {alsoletter={\#, \&, $},
  mathescape=true,
  basicstyle=\small,
  escapechar=\@,
  boxpos=c,
  morekeywords= {and, axioms, axiomatization, Skip, Stop, type_synonym, theorem, lemma, apply, by, constdefs, definition, where,
    infixl, types, consts, primrec, have, from, show, let,  proof, qed, is,
    sorry, with, assume, fix, thus, hence, datatype, if, then, else, in, case,
    of
  },
  emph={[2] real, int, char, bool, string, self, boolean, Ass, Send, Rec, Seq, Cond, Pref,
                Join, Meet, Par, Rep, Cont, TOut, Inrp},
  emphstyle={[2]\it},
literate=
  {->}{{$\rightarrow$}}2
  {=>}{{$\Rightarrow$}}2
  {-->}{{$\Rightarrow$}}2
  {\\forall}{{$\forall$}}2
  {\\exist}{{$\exists$}}2
  {\\exists}{{$\exists$}}2
  {AND}{{$\& \&$}}3
  {ALL}{{$\forall$}}2
  {EX}{{$\exists$}}2
  {\%}{{$\lambda$}}1
  {\\\/}{{$\sqcap$}}2
  {|-}{{$\vdash$}}2
  {==}{{$\equiv$}}2
  {==>}{{$\Rightarrow$}}2
  {~}{{$\neg$}}1
  {~=}{{$\neq$}}2
  {:=}{{\tt :=}}2
  {;}{{\tt ;}}1
  {$}{{\do}}1
}

\lstnewenvironment{isaenv} {
  \lstset{extendedchars=true,columns=flexible}
  \lstset{language=isa,deletestring=[b]', basicstyle=\footnotesize}%
} {}
\newcommand{\isa}[1]{{\lstinline[language=isa, basicstyle={\footnotesize \tt} ]+#1+}}

\title{A Theorem Prover for Quantum Hoare Logic and Its Applications}
\author{Tao Liu \inst{1}, Yangjia Li \inst{1}, Shuling Wang \inst{1}, Mingsheng Ying \inst{2} and Naijun Zhan \inst{1}}
\institute{State Key Lab. of Computer Science, Institute of Software, CAS\\
\email{\{liut,yangjia,wangsl,znj\}@ios.ac.cn} \and
  Center for Quantum Computation and Intelligence Systems, Uni. of Tech. Syndney \\
\email{Mingsheng.Ying@uts.edu.au}
}
\begin{document}
\maketitle
\begin{abstract}
Quantum Hoare Logic (QHL) was introduced in \cite{Ying12} to specify and reason about
  quantum programs. In this paper, we implement a theorem prover for QHL  based on
   Isabelle/HOL. By applying the theorem prover,  verifying a quantum program against
   a specification is transformed equivalently into an order relation between matrices. Due to the limitation of Isabelle/HOL, the calculation of the order relation is  solved by calling an outside oracle written in  Python.
To the best of our knowledge, this is the first theorem prover for quantum programs. To demonstrate its power,
the correctness of two well-known quantum algorithms, i.e., Grover Quantum Search and Quantum Phase Estimation (the key step in Shor's quantum algorithm of factoring in polynomial time) are
proved using the theorem prover.  These are the first mechanized proofs for both of them.

\end{abstract}
\section{Introduction}
Due to the rapid progress in quantum technology in the recent years, it is predicted practical quantum computer hardware can be built within 10-15 years. On the other hand, quite intensive research on quantum programming has been conducted in the last decade \cite{BCS01,Omer03,SZ00,Selinger04\oomit{,YF11}}, as surveyed in \cite{Gay06,Selinger04b\oomit{,YFD+12}}. In particular, several quantum programming languages have been defined and their compilers have been implemented, including Quipper \cite{Green13}, Scaffold \cite{Ali15} and Microsoft's LIQUi$|>$ \cite{Wecker}. Several techniques, including model-checking, and tools have been developed for verification and analysis of quantum programs and quantum cryptographic protocols \cite{GNP08,AGN13,AGN14,YYF+13}. Along this line, a few program logics and proof systems have been proposed to reason about the correctness of quantum programs \cite{BS06,BJ04,CMS06,FDJ+07,Ying12}. A recent result of this direction is the \emph{Quantum Hoare Logic} (QHL)~\cite{Ying12}. It extends to the sequential quantum programs the \emph{Floyd-Hoare-Naur} inductive assertion methods - the dominant approach of proving the correctness of classical programs. QHL is proved to be (relatively) complete for both partial correctness and total correctness of quantum programs.

Like in the verification of classical programs, it is absolutely necessary to provide tool supports in the verification of quantum programs. This motivates us to implement a theorem prover for QHL to assist the user to mechanize the proofs of quantum programs with
QHL. The contribution of the paper is two-fold:\begin{itemize}
\item We implement a theorem prover for reasoning about quantum programs based on QHL. The theorem prover is composed of two parts: a verification condition generator (VCG) in Isabelle/HOL, and an external oracle in Python aided for performing computations on matrices. The theoretical foundation of the VCG is the proof system of QHL for reasoning about quantum programs. In the implementation, we focus on the case of partial correctness for QHL, as the termination of most existing quantum algorithms can be proved by the specific techniques, such as~\cite{LYY14}.

\ \ \ In detail, to implement the VCG, we embed the whole framework of QHL including the syntax, semantics and the proof system into Isabelle/HOL. In QHL, a specification for a quantum program $S$ is written as $\{Pre\} S \{Post\}$, where precondition $Pre$ and postcondition $Post$ are Hermitian matrices (called observables in quantum physics). By applying the VCG, it is equivalently transformed to an L\"{owner} order relation $Pre \sqsubseteq Pre'$ (i.e. $Pre'-Pre$ is a positive semi-definite matrix) for some $Pre'$. We implement an oracle in Python for deciding such order relations between matrices.  By invoking two special packages \emph{Numpy} and \emph{Sympy} to do matrix computation, the oracle is able to check whether $Pre \sqsubseteq Pre'$ holds exactly and  efficiently. Finally, the result is sent back to Isabelle/HOL and   the proof is  completed.
\item To demonstrate the power of the theorem prover, the correctness of two well known quantum algorithms, i.e., \emph{Grover Quantum Search} and \emph{Quantum Phase Estimation} are proved using the prover. Grover Quantum Search provides a quadratic speedup for searching in a unsorted database. Quantum Phase Estimation is the key step in Shor's quantum algorithm of factoring in polynomial time, and it is also a crucial step in Harrow-Hassidim-Lloyd quantum algorithm for linear system of equations. This is the first time that the correctness of these two quantum algorithms is proved in a theorem prover.
\end{itemize}

\subsection{Related work}
In the literature, quite a few verification tools have been developed for quantum systems, but almost
all of them target checking quantum communication protocols. For
example, Nagarajan and Gay \cite{NG02} modeled the BB84 protocol~\cite{BB84} in the classical CCS and verified its correctness by using the Concurrency Workbench of the New Century. Taking the probabilism arising from quantum
measurements into account, the probabilistic model checker PRISM \cite{KNP04} is further used~\cite{GNP05} to verify the correctness of several quantum protocols including BB84. Furthermore, Gay et al~\cite{GNP08} and Papanikolaou~\cite{Papanikolaou08}
developed an automatic tool Quantum Model-Checker (QMC). QMC uses stabilizer formalism
\cite{Gottesman97} for the modeling of systems, and the properties to be checked
by QMC are expressed in QCTL (Quantum Computation Tree Logic) defined in \cite{BCM+07,BCM08}. For a systematic exposition of the results in this research
line, see \cite{GNP10} and \cite{Papanikolaou08}. More recently, Ardeshir-Larijani et al. \cite{AGN13,AGN14} presented a tool for verification of quantum protocols through equivalence checking. Boender et al. \cite{BKN15} proposed a framework for specifying and verifying quantum protocols using Coq.

However, all of the tools mentioned above are inapplicable to verification for general quantum programs, as they are restricted to some special classes of  quantum systems, e.g. those in the stabilizer formalisms. Recently, several specific techniques have been proposed to algorithmically check properties of quantum programs. In \cite{YYF+13}, the Sharir-Pnueli-Hart method for proving probabilistic programs~\cite{SPH84} has been generalised to quantum programs by exploiting the Schr\"{o}dinger-Heisenberg
duality between quantum states and observables. Termination analysis of nondeterministic and
concurrent quantum programs~\cite{LYY14\oomit{,YY12}} was carried out based on reachability analysis of
quantum Markov decision processes~\cite{LY14,YFY+13}. But up to now no tools have been developed to implement these techniques.

The mechanized theorem proving has made a significant progress in the certified verification of classical programs since 21st century. One of the most well-known work is the CompCert C verified compiler back-end~\cite{Leroy}, which is developed with the proof assistant Coq  and is robust enough for building critical embedded softwares. Another work is the L4.verified project~\cite{FultonMQVP15}, which establishes a mechanized proof of the correctness of the seL4 microkernel using the proof assistant Isabelle/HOL. Also using  Isabelle/HOL, the Verisoft project~\cite{Alkassar} developed  the formal verification from the application layer over the system level software in C language,  down to the assembler and hardware. Recently, the verification of hybrid systems via interactive theorem proving has been studied. The theorem prover KeYmaera X~\cite{FultonMQVP15} provides a mathematical correctness proof for hybrid programs consisting both the control program and the physical dynamics together. In our previous work~\cite{ZWZ13}, we have developed an interactive theorem prover for reasoning about hybrid CSP~\cite{He94} processes within Isabelle/HOL. We expect that theorem provers will play a crucial role in quantum computing, as they did in classical computing, with this paper as one of the first steps.

\section{Preliminary}
In this section, we briefly recall the basic concepts and results of quantum Hoare logic (QHL). We only introduce the proof system for partial correctness, since the \emph{rule loop total} for total correctness is not implemented in our prover. The complete version of QHL can be found in \cite{Ying12}.

Quantum programs are formally defined by the following grammar:
$$S::={\bf skip}\mid q:=0\mid \overline{q}:=U\overline{q}\mid S_1;S_2\mid{\bf measure}\ M[\overline{q}]:\overline{S}\mid{\bf while}\ M[\overline{q}]=1\ {\bf do}\ S$$
where\begin{itemize}
\item $q$ is a quantum variable which should be one of the two types {\bf Boolean} and {\bf integer}, and $\overline{q}$ is a finite sequence of distinct quantum variables, called a \emph{quantum register};
\item $U$ is a unitary operator on the state (Hilbert) space $\mathcal{H}_{\overline{q}}$ of $\overline{q}$;
\item In ${\bf measure}\ M[\overline{q}]:\overline{S}$, $M=\{M_m\}$ is a quantum measurement on $\mathcal{H}_{\overline{q}}$, and $\overline{S}=\{S_m\}$ is a set of quantum programs such that each possible outcome $m$ of the measurement corresponds to $S_m$;
\item $M=\{M_0,M_1\}$ in ${\bf while}\ M[\overline{q}]=1\ {\bf do}\ S$ is a yes-no measurement on $\overline{q}$.
\end{itemize}
In fact, such quantum programs can be regarded as quantum extensions of classical {\bf while}-programs. The skip statement does nothing but terminates, which is as the same as in the classical case. The initialization statement sets quantum variable $q$ to the basis state $\ket{0}$ and the statement of unitary transformation changes the state of $\overline{q}$ according to $U$, so both of them are quantum counterparts of the assignment statement in classical programming languages. The sequential composition is similar to its classical counterpart. The measurement statement is a quantum generalisation of the classical case statement: ${\bf if}\ (\Box m\cdot b_m\rightarrow S_m)\ {\bf fi}$. The loop statement is also a quantum generalisation of classical loop ``{\bf while} $b$ {\bf do} $S$''. Formally, the denotational semantics of a quantum program $S$ is defined as a super-operator  $\llbracket S\rrbracket(\cdot)$, such that if the input state is $\rho$ (a density matrix), then $\llbracket S\rrbracket(\rho)$ is the sum of output states (partial density operators) of terminating computations of $S$.

The correctness of a quantum program $S$ is expressed by a quantum extension of the Hoare triple $\{P\}S\{Q\}$, where the precondition $P$ and postcondition $Q$ are formalized by Hermitian matrices or physical observables between zero and identity matrices, called quantum predicates \cite{DP06}. The semantics of the partial correctness formulas is defined as follows: $$\models_{\textit{par}}\{P\}S\{Q\} \mbox{   iff   } \textit{tr}(P\rho)\leq \textit{tr}(Q\llbracket S\rrbracket(\rho))+\textit{tr}(\rho)-\textit{tr}(\llbracket S\rrbracket(\rho))$$ for all quantum states $\rho$. Here $\llbracket S\rrbracket(\cdot)$ represents the denotational semantics of $S$. The semantics of the total correctness formulas is defined similarly: $$\models_{\textit{tot}}\{P\}S\{Q\} \mbox{   iff   } \textit{tr}(P\rho)\leq \textit{tr}(Q\llbracket S\rrbracket(\rho))$$ for all quantum states $\rho$. We note that they become the same when the quantum program $S$ is terminating, i.e. $\textit{tr}(\llbracket S\rrbracket(\rho))=\textit{tr}(\rho)$ for all quantum states $\rho$.

The proof system $\textit{qPD}$ for partial correctness of quantum programs is given in Fig.~\ref{fig:qpd}. The soundness and (relative) completeness of $\textit{qPD}$ has been proved in \cite{Ying12}.
\begin{theorem}
The proof system $\textit{qPD}$ is sound and (relative) complete, for partial correctness of quantum programs.
\end{theorem}

The concept of \emph{quantum weakest liberal precondition} is very helpful for implementation of $\textit{qPD}$. Formally, the weakest liberal precondition of a quantum program $S$ with respect to a quantum predicate $P$, denoted by $wlp.S.P$, is the weakest quantum predicate $Q$ satisfying $\models_{\textit{par}}\{Q\}S\{P\}$. It follows immediately from the semantics of correctness formula that $wlp.S.P=I-\llbracket S\rrbracket^*(I-P)$, where $\llbracket S\rrbracket^*(\cdot)=\sum_i E_i^\dagger\cdot E_i$ represents the Schr\"{o}dinger-Heisenberg dual of $\llbracket S\rrbracket(\cdot)=\sum_i E_i\cdot E_i^\dagger$ (Kraus form of superoperators). Computation of the weakest liberal preconditions can be achieved by the following rules: \\
\begin{tabular}{ll}
 (a) & $\textit{wlp}.\textbf{skip}.P=P$; \\[2mm]
 (b) & If $\textit{type}(q)=\textbf{Boolean}$, then
     $\textit{wlp}.(q:=0).P={\ket {0}}_q \bra 0 P {\ket 0}_q {\bra 0} + {\ket 1}_q {\bra 0} P {\ket 0}_q {\bra 1}$, \\
    &    if $\textit{type}(q)=\textbf{integer}$, then
      $\textit{wlp}.(q:=0).P=\sum\limits_{n=-\infty}^{\infty}  {\ket n}_q \bra 0  P {\ket 0}_q {\bra n}  \rbrace$;  \\[2mm]
(c) & $\textit{wlp}.(\overline{q}:=U\overline{q}).P=U^{\dag} \textit{PU}$; \\[2mm]
(d) & $\textit{wlp}.(S_1;S_2).P=\textit{wlp}.S_1.(\textit{wlp}.S_2.P)$ ; \\[2mm]
 (e) & $\textit{wlp}.(\textbf{measure}\ M[\overline{q}]:\ \overline{S}).P=\sum_m {M_m}^{\dag} (\textit{wlp}.S_m.P) M_m$; \\[2mm]
 (f) & $\textit{wlp}.(\textbf{while}\ M[\overline{q}]=1\ \textbf{do}\ S).P=\bigsqcap_{n=0}^{\infty} P_n$ , where \\[3mm]
  & \hspace*{2cm}  $\left \{
 \begin {array}{lcl}
  P_0=I, \\
  P_{n+1}=M_0^{\dag} PM_0+M_1^{\dag} (wlp.S.P_n) M_1 \mbox{ for all } \ n\geq 0
 \end{array}
 \right.
 $\\[3mm]
 & $I$ is the identity operator in the state (Hilbert) space of all quantum variables,\\ & and $\bigsqcap$ stands for the greatest lower bound of quantum predicates according to\\ &the L\"{o}wner partial order $\sqsubseteq$.
\end{tabular}

\begin{figure}[!t]
\centering
\fbox{
\begin{tabular}{ll}
\textbf{(Skip)} &
$ \lbrace P \rbrace \ \textbf{skip} \ \lbrace P \rbrace $
\\[5mm]
\textbf{(Init)} & (1) If $\textit{type}(q)=\textbf{Boolean}$, then\\[1mm]
     & \qquad \qquad$\lbrace {\ket {0}}_q \bra 0 P {\ket 0}_q {\bra 0} + {\ket 1}_q {\bra 0} P {\ket 0}_q {\bra 1}  \rbrace \  q:=0 \ \lbrace P \rbrace$\\[1mm]
    &  (2) If $\textit{type}(q)=\textbf{integer}$, then\\[1mm]
    &  \qquad \qquad$\lbrace  \sum\limits_{n=-\infty}^{\infty}  {\ket n}_q \bra 0  P
{\ket 0}_q {\bra n}  \rbrace   \   q:=0     \lbrace P \rbrace$
\\[5mm]
\textbf{(UT)} &
$\lbrace U^{\dag} PU \rbrace \ \overline{q}:=U\overline{q} \ \lbrace P \rbrace$\\[5mm]
\textbf{(Seq)} &
  $\frac{\displaystyle \lbrace P \rbrace \ S_1 \ \lbrace Q \rbrace \qquad \lbrace Q \rbrace \ S_2 \ \lbrace R \rbrace }{\displaystyle \lbrace P \rbrace \ S_1;S_2 \ \lbrace R \rbrace}$ \\[5mm]
\textbf{(Mea)} &
   $\frac{\displaystyle \lbrace P_m \rbrace \ S_m \ \lbrace Q \rbrace\ \mbox{for all} \ m}{\displaystyle \lbrace \sum_m {M_m}^{\dag} P_m M_m \rbrace \  \textbf{measure}\ M[\overline{q}]:\ \overline{S} \ \lbrace Q \rbrace }$ \\[5mm]
\textbf{(Loop)} &
   $\frac{\displaystyle \lbrace Q\rbrace \ S  \  \lbrace M_0^{\dag} P M_0 + M_1^{\dag} Q M_1 \rbrace}
{\displaystyle \lbrace  M_0^{\dag} PM_0+M_1^{\dag} Q M_1 \rbrace \  \textbf{while}\ M[\overline{q}]=1\ \textbf{do}\ S \ \lbrace P\rbrace}$ \\[5mm]
\textbf{(Order)} \quad \quad  \quad \quad &
  $\frac{\displaystyle P\sqsubseteq P^{\prime} \quad \lbrace P^{\prime} \rbrace \ S \ \lbrace Q^{\prime} \rbrace \quad  Q^{\prime} \sqsubseteq Q}{\displaystyle \lbrace P \rbrace \ S \ \lbrace Q\rbrace}$
\end{tabular} }
\caption{Proof system $\textit{qPD}$ of partial correctness}
\label{fig:qpd}
\end{figure}

\textbf{The idea of proving correctness of a quantum program}: To prove a correctness formula $\{P\}S\{Q\}$, we start from the postcondition $Q$ (a Hermitian matrix) and try to compute the weakest liberal precondition (also a Hermitian matrix) for each statement $S_i$ within $S$, which can be easily done according to the rules provided above if $S_i$ is not the loop statement. However, when $S_i$ is a loop statement, the computation rule of $\textit{wlp}$ cannot be directly followed. In this case, we will find a invariant $R$ of the loop body, then the precondition can be chosen as $M_0^{\dag} Q M_0+M_1^{\dag} R M_1$ by the rule \textbf(Loop) of $qPD$. At last, we obtain a precondition $P^{\prime}$ (a Hermitian matrix) at the beginning of the program, and it suffices to prove that $P\sqsubseteq P^{\prime}$. As one can imagine, the computation in such a proof is much harder than the proof of a classical program due to the involved matrix manipulations.

\section{Implementation }
In this section, we first give an overview of the theorem prover for QHL, and then
explain the details of its implementation including the mechanization of QHL in Isabelle/HOL and the external oracle in Python for deciding the order of matrices.

\newcommand{\commentwsl}[1]{{\color{red} \emph{#1}}}


\subsection{The Architecture in a Nutshell}
The theoretical foundation of the theorem prover  is the proof system of QHL introduced in the previous section. It is composed of two parts: a verification condition generator for quantum programs in Isabelle/HOL at the frontend, and an external oracle in Python aided for calculation at the backend. Unlike the theorem prover for classical Hoare logic, QHL prover tends to do a lot of computations, especially matrix operations. So the oracle is a calculation module out of Isabelle/HOL designed for performing all the operations over matrices.

First of all, the prover reads a quantum specification, including a quantum program $S$ and two quantum predicates $P$ and $Q$ (Hermitian operators) for describing the precondition (i.e. input state) and the postcondition (i.e. output state) respectively. The specification is encoded in the syntax we have defined in Isabelle/HOL, and especially, the operators (matrices) in a program are identified  by some symbols, like $P, Q, R$ and  whatever you want, while the concrete matrices will be recorded in a .txt file outside the Isabelle/HOL to be used in the calculation later. By applying the verification condition generator, we obtain an order relation $P \sqsubseteq P'$, where $P$ is the given precondition, and $P'$ is the  precondition of $S$ with respect to the postcondition $Q$ deduced based on QHL. All the left is to decide whether  $P \sqsubseteq P'$ holds (i.e. $P'-P$ is positive semi-definite), which is done by the  oracle  developed outside Isabelle/HOL. The oracle receives a string parsed in ML that corresponds to the order relation, and  a file named ``param.txt'' providing the concrete matrices of the quantum program corresponding to the abstract matrix symbols in Isabelle/HOL  as input, and returns the result ``Yes'' or ``No'' as output. This result will then be passed back to Isabelle/HOL to complete the proving.

In a nutshell, the main idea is to record the process of deducing the order relation in Isabelle/HOL and  compute the relation all at one time by an oracle outside. This is a bit like  what the Apache Spark, a fast and general-purpose cluster computing system, does.

\subsection{Mechanization in Isabelle/HOL}
Isabelle/HOL is a proof assistant for Higher-Order Logic (HOL). It supports functional modeling of systems and formulation of properties
by providing datatypes, functions, and formulas. Isabelle/HOL enables the proof by induction directly. Besudes a set of rules and  methods for classical reasoning, it also includes some high-level proof tactics, for example, the tool \isa{sledgehammer}, which is a certified integration of third-party automated theorem
provers  and SMT solvers. To mechanize  QHL in Isabelle/HOL, we need to embed the whole logic framework including the syntax, semantics and proof system.

The matrices and matrix calculations are extensively involved in the deductive reasoning of QHL.  Due to the limited support for operations over matrices in Isabelle/HOL, we declare a new type \isa{Mat} to represent matrices and define its meaning by a set of axioms. This is discouraged in Isabelle/HOL, but considering the tradeoff between proof soundness and efficiency, we allow two kinds of axioms:
\begin{itemize}
  \item the basic properties of matrices, such as  $(A+B)C = AC +BC$, $tr(AB) = tr(BA)$, etc;
  \item the Knaster-Tarski fixpoint theorem.

\end{itemize}

Based on type \isa{Mat},  the syntax of quantum programs is represented as a datatype \isa{com} as follows:
\begin{isaenv}
   datatype com = SKIP | Init (nat list) nat | Utrans Mat nat
                      |  com; com  | Cond  (Mat * com * Mat) list
                      | While  Mat Mat com Mat | While_n Mat Mat com Mat nat
\end{isaenv}
besides the last one, each of which corresponds to the respective quantum construct presented in Sec. 2. \isa{While_n} is defined for the purpose of the verification of the while loop \isa{While}. For simplicity, for each variable in the quantum program, we assign  a unique number to represent it in the implementation.

\textbf{Encoding}: For initialization  $q_i:=0$, $q_i$ can be of type {\bf Boolean} or {\bf integer}, in the implementation, we use the size of the state space of the  variable  to mark the different type. For instance, if $q_i$ is {\bf Boolean}, then  $q_i:=0$ is encoded as \isa{Init [1, 2] k}, \isa{k} for the number corresponding to the variable. $\overline{q}:=U\overline{q}$ is encoded as \isa{Utrans U n}, where \isa{n} denotes the number corresponding to the register $\overline{q}$. $S_1;S_2$ is encoded as \isa{S_1; S_2}, for which \isa{S_i} for $i=1,2$ corresponds to the encoding of $S_i$ resp. ${\bf measure}\ M[\overline{q}]:\overline{S}$ is encoded as
$$\isa{Cond [(M_0, S_0, Q_0), ..., (M_n-1, S_n-1, Q_n-1)]},$$
 where $n$ denotes the length of $\overline{S}$, also of $\overline{q}$, \isa{Q_i} for \isa{i = 0, ..., n-1} is introduced for the purpose of verification. At last, the loop ${\bf while}\ M[\overline{q}]=1\ {\bf do}\ S$ is encoded as \isa{While M_0 M_1 S Q}, where \isa{[M_0, M_1]} represents $M$, \isa {Q} is the loop invariant introduced for verification purpose.

%

 Given a statement \isa{S} and a matrix \isa{P}, the recursive function \isa{denoFun S P} defines the denotational semantics of $S$, i.e. the encoding of  $\llbracket S\rrbracket(P)$ exactly. Meanwhile, given a statement \isa{S} and a matrix \isa{Q}, the function
 \isa{wlp Q S} defines the weakest liberal precondition of \isa{S} with respect to \isa{Q}, i.e. the encoding of $ wlp.S.Q$ exactly. Based on the semantics,
  given a statement \isa{S} and two matrices \isa{P} and \isa{Q},  \isa{valid P S Q} is implemented to represent a valid Hoare triple.

\textbf{Soundness}: In~\cite{Ying12}, two forms of proof systems of QHL are presented, one is structural as presented in Fig.~\ref{fig:qpd}, and the other one is based on weakest liberal precondition. We have proved that both are sound  with respect to the denotational semantics. As an illustration, we have proved the following lemma stating the soundness of the rule for unitary transformation:
  \begin{isaenv}
  lemma Utrans: valid  (mat_mult (mat_mult (dag U) P) U)  (Utrans U n) P
  \end{isaenv}
  and for the weakest liberal precondition, we have proved a lemma  stating the validity of all the cases as a whole:
\begin{isaenv}
  lemma WLPsound: $\forall$ Q. valid (wlp Q S) S Q
\end{isaenv}


Besides the lemmas corresponding to the inference rules specific to each quantum construct, we also prove the following order rule applicable for all constructs:
\begin{isaenv}
  lemma Order: less P P1 ==> valid P1 S Q1 ==> less Q1 Q ==> valid P S Q
\end{isaenv}where \isa{less P P1} represents $P \sqsubseteq P1$, i.e. $P1 - P$ is positive.
At last, we get the rule $ord\_wlp$ which combines the rules \isa{Order} and  \isa{WLPsound} above:
\begin{isaenv}
  lemma ord_wlp: less P (wlp Q S) ==> valid P S Q
\end{isaenv}
All the above lemmas constitute a verification condition generator of QHL.

We can  intuitively understand the process of proof as follows:  if user wants to prove \isa{valid pre example post}, where \isa{pre} and \isa{post} are the pre-/post-matrices, and \isa{example} is the quantum program, then by applying the verification condition
generator, \isa{valid pre example post} is finally reduced to  an order relation \isa{less pre pre'}, where \isa{pre'} represents the deduced precondition of \isa{example} with respect to \isa{post}. In fact, if \isa{example} does not contain the while statement, \isa{pre'} will be exactly the weakest liberal precondition of \isa{example} with respect to \isa{post}, i.e. \isa{wlp post example}. However, the precondition of a while statement with respect to a given postcondition is defined by using the  annotated loop invariant, thus it is not the weakest. Finally, the relation will then be proved by calling an external oracle, which will be explained subsequently.

\subsection{Oracle in Python}
Python \cite{python} is an object-oriented programming language, which provides an interpreter so that
prototyping a system to be developed is supported by the dynamic development environment.
Thus, Python supports \emph{agile software engineering}, and becomes very popular now.
 In addition, scientific computing can be easily coped with by Python with
   the two packages \emph{Numpy} and \emph{Sympy}. \emph{Numpy} is used for numerical computation, which contains most of the operations over matrices, and
   \emph{Sympy} is used for conducting  symbolic computation. Here we combine the symbolic computation of \emph{Sympy} and matrix operations of \emph{Numpy} to conduct symbolic matrix computation.

   As mentioned above, the lemma to be proved \isa{valid pre example post} is reduced to an order relation \isa{less pre pre'}. Thus, the left is to check whether \isa{pre'-pre}  is positive.
   The precondition \isa{pre'} is deduced based on the inference system of QHL, and it is not a real matrix but just a long term in Isabelle/HOL representing a matrix. The term must be transformed into a string first and then be passed to Python, which is done by a parser module written in ML.
   The computing module that is  developed with \emph{Numpy} and \emph{Sympy} packages in Python then receives the  string, and a file named ``param.txt" providing the concrete matrices of the quantum program that correspond to those matrix symbols in Isabelle/HOL,  as input, and returns the result ``Yes'' or ``No'' as output, which is passed back to Isabelle/HOL to complete the proof.

  When computing the matrix corresponding to the term \isa{pre'} by using the oracle, we must pay attention to a key point as follows. The  calculation is in the Hilbert space, and the state space of a quantum system is the tensor product of the state spaces of the variables which we also call components.
Suppose there is a quantum program where there are $n$ {\bf boolean} variables $q_0, q_1\cdots, q_{n-1}$ , the state space is
{ \[\begin{array}{l}
\mathcal{H}=\mathop{\bigotimes}\limits_{i=0}^{n-1} \mathcal {H}_{q_i}
\end{array} \] } If there is a program statement, e.g. $q_i=Uq_i$ , and its post condition is $Q$, then we can get the weakest libral precondition
{ \[\begin{array}{l}
P=\left(\mathop{\bigotimes}\limits_{k=0}^{i-1} I_{q_k} \otimes U\dag    \otimes  \mathop{\bigotimes}\limits_{k=i+1}^{n-1} I_{q_k}\right) Q \left(\mathop{\bigotimes}\limits_{k=0}^{i-1} I_{q_k} \otimes U   \otimes  \mathop{\bigotimes}\limits_{k=i+1}^{n-1} I_{q_k} \right)
\end{array} \] } So when recording the calculation process, we must consider the order of variables which is quite different from that of classical program. This is also the main reason why we use numbers to represent the quantum variables in the syntax encoding.

\section{Applications}
In this section, in order to demonstrate the power of the QHL prover, we prove the correctness of two well-known quantum algorithms, i.e., \emph{Grover Quantum Search} and  \emph{Quantum Phase Estimation} with the latter being the key step in Shor's quantum algorithm of factoring in polynomial time and a crucial step in Harrow-Hassidim-Lloyd quantum algorithm for linear systems of equations.
 To the best of our knowledge, these are the first mechanical proofs for both of them.

\subsection{Grover Quantum Algorithm}
The Quantum Grover Algorithm is presented in Algorithm~\ref{alg:grover}, for which especially the right part gives the encoding of the algorithm in Isabelle/HOL. The search problem can be explained as follows: Suppose the search space consists of N elements,
indexed by numbers $0, 1, \cdots, N-1$, and there are exactly M solutions among the N elements for $1 \leq M \leq \frac{N}{2}$. In Algorithm~\ref{alg:grover}, we assume $N=2^n$, so  the elements of the search space can be stored in n qubits, that is $q_0$, $q_1$, $\cdots$, $q_{n-1}$.  The $O$ in the loop body $D$ is an oracle, that can be considered as a black box
with the ability to recognize solutions to the search problem; and Ph is an operator preforming a condition phase shift.  H is a Hadamard operator. We write $ \overline{q}$ for the quantum register consisting of $q_0$, $q_1$, $\cdots$, $q_{n-1}$. The variable $r$  is introduced to count the number of iterations
of the Grover operator. The measurement $M=\lbrace M_0, M_1\rbrace$ in the loop is given as follows:
{ \[\begin{array}{l}
M_0=\sum\limits_{k \geq R} \ket{k}_r \bra{k},\ \ \ \ \ \ \ \
M_1=\sum\limits_{k < R} \ket{k}_r \bra{k},
\end{array} \] }in which $R$ is the number of iterations of the algorithm and its calculation is omitted here.
The loop body is given in Sub-algorithm~\ref{alg:groverBody}. In the measurement statement, $N=\lbrace N_t=\ket {t} \bra {t}:t \in {\lbrace 0, 1 \rbrace}^n  \rbrace$
is the measurement in the computational basis of n qubits, and all elements in $\overline{S}$ are \textbf{skip}, that is, $\overline S =\lbrace S_t\equiv \textbf{Skip}: t \in {\lbrace 0, 1 \rbrace}^n \rbrace$.

\begin{multicols}{2} 
\begin{algorithm}[H]
 $q_0:=0$; \\
 $q_1:=0$ ; \\
 \quad ... ...  \\
 $q_{n-1}:=0$; \\
 $q:=0$; \\
 $r:=0$; \\
 $q:=\sigma_xq$; \\
 $q_0:=Hq_0$;  \\
 $q_1:=Hq_1$; \\
\quad   ... ...  \\
 $q_{n-1}:=Hq_{n-1}$; \\
 $q:=Hq$;  \\
 \textbf{while}\ $M[r]=1$\ \textbf{do} \  D;  \\
 \textbf{measure} $N[q_0,q_1,...,q_{n-1}]:\bar{S}$ \; \\
  \\
 \caption{Quantum Grover Search} \label{alg:grover}
\end{algorithm}
\columnbreak
\hfill
\begin{minipage}[b]{0.48\textwidth}
\centering
\vspace{-0.0cm}
\begin{isaenv}

Init [1,2] 0;
Init [1,2] 1;
 ... ...
Init [1,2] n-1;
Init [1,2] n;
Init [1,2] n+1;
Utrans Delta n;
Utrans H 0;
Utrans H 1;
 ... ...
Utrans H n-1;
Utrans H n;
While M0 M1 D Q;
Cond [(N0,SKIP,p0), ...,
      (N_n-1,SKIP,p_n-1)]
\end{isaenv}
\vspace{-0.0cm}
\end{minipage}
\end{multicols}

\medskip

\floatname{algorithm}{Sub-algorithm}
\setcounter{algorithm}{0}

\begin{multicols}{2} 
\begin{algorithm}[H]
 $q_0,q_1,...,q_{n-1}:=O\ q_0,q_1,...,q_{n-1}$;\\
 $q_0:=Hq_0$;\\
 $q_1:=Hq_1$;\\
 \quad ... ...  \\
 $q_{n-1}:=Hq_{n-1}$;\\
 $q_0,q_1,...,q_{n-1}:=Ph\ q_0,q_1,...,q_{n-1}$;\\
 $q_0:=Hq_0$;\\
 $q_1:=Hq_1$;\\
 \quad ... ...  \\
 $q_{n-1}:=Hq_{n-1}$;\\
  $r:=U_{+1}r$  \\
   \\
 \caption{Loop body D} \label{alg:groverBody}
\end{algorithm}
\columnbreak
\begin{minipage}[b]{0.49\textwidth}
\centering
\vspace{-0.2cm}
\begin{isaenv}


Utrans O n-1...10
Utrans H 0;
Utrans H 1;
 ... ...
Utrans H n-1;
Utrans Ph n-1...10
Utrans H 0;
Utrans H 1;
 ... ...
Utrans H n-1;
Utrans U n+1

\end{isaenv}
\vspace{-0.5cm}
\end{minipage}
\end{multicols} 

\floatname{algorithm}{Algorithm}

The correctness of the program can be stated as
 $$\models \lbrace p_{\textit{succ}} \ I \rbrace \ \  \textit{Grover} \ \  \lbrace P\rbrace,$$  where
{ \[\begin{array}{lcrlcr}
I&=&\mathop{\bigotimes}\limits_{i=0}^{n-1} I_{q_i} \otimes I_q \otimes I_r, &
P & = & \left(\sum\limits_{t\ solution}  \ket{t}_{\overline q} \bra {t} \right) \otimes I_q \otimes I_r.
\end{array} \] }
 $p_{\textit{succ}}$ is the probability with which the algorithm will achieve a success.
$I_{q_i}(t=0, 1, \cdots, n-1)$ and $I_q$ are the identity operators on  $\mathcal{H}_2$ (the $2$-dimensional Hilbert space, i.e. the state space of qubit or a quantum variable of type \textbf{Boolean}); and $I_r$ is the identity operator on $\mathcal{H}_{\infty}$ (the infinite-dimensional Hilbert space, i.e. the state space of a quantum variable of type \textbf{integer}).


To avoid too complicated calculation, we show the case of $N=4$ (i.e. $n=2$) and $M=1$. The more general cases  can be handled similarly, but the computational complexity may increase exponentially  with the growth of $n$, and moreover, the input of matrices would  be quite troublesome. This is the bottleneck that we verify quantum programs using classical computers. In the case for $N=4$, there is a unique solution, say $s$, so we can get $P={\ket {s}}_{\overline q} \bra {s}  \otimes I_q \otimes I_r$. It is not difficult to prove  $p_{\textit{succ}}=1$. Thus,  we need to prove \isa{  valid  \{  I  \}  \ Grover  \{ P \}}, which is equivalently transformed to \isa{I} $\sqsubseteq$ \isa{ P'} by applying \isa{vcg}, where \isa{P'} is the deduced precondition of \isa{Grover} with respect to \isa{P}. In detail, the order relation \isa{I} $\sqsubseteq$ \isa{ P'} corresponds to the following term in Isabelle/HOL:

\begin{isaenv}
basic.less I
     (matsum q0 0
       (matsum q1 (Suc 0)
         (matsum q 2
           (matsum r 3
             (matUtrans Delta 2
               (matUtrans H 0
                 (matUtrans H (Suc 0)
                   (matUtrans H 2
                     (fixpoint_wlp M0 M1 C Q
                       (mat_add (mat_mult (mat_mult (dag N0) P) N0)
                         (mat_add (mat_mult (mat_mult (dag N1) P) N1)
                           (mat_add (mat_mult (mat_mult (dag N2) P) N2)
                             (mat_add (mat_mult (mat_mult (dag N3) P) N3) zero)
                            ))))))))))))
\end{isaenv}
\oomit{\isa{" basic.less I
    (matsum q0 0
       (matsum q1 (Suc 0)
         (matsum q 2
           (matsum r 3  }\\
            \isa{  (matUtrans Delta 2
               (matUtrans H 0
                 (matUtrans H (Suc 0)
                   (matUtrans H 2    }\\
                   \isa{  (fixpoint_wlp M0 M1 C Q
                     (mat_add (mat_mult (mat_mult (dag N0) P) N0)}\\
                         \isa{  (mat_add (mat_mult (mat_mult (dag N1) P) N1)
                           (mat_add (mat_mult (mat_mult} \\ \isa {(dag N2) P) N2) (mat_add (mat_mult (mat_mult (dag N3) P) N3) zero)))))))))}\\ \isa{))))  "}. }
where
\begin{eqnarray*}
\isa{matsum q i P} & = & A\dag  P A, \\
 \isa{matUtrans H i P} &=& B\dag  P B ,  \\
\isa{fixpoint_wlp M0,M1,C,Q P} &=& M0^{\dag} P M0 + M1^{\dag} Q M1, \mbox{ in which}\\
A&= &( I_{q_0} \otimes I_{q_1} \otimes   (\ket 0 \bra 0+\ket 0 \bra 1)  \mathop{\otimes} I_{q_r} )\\
     B&=&(I_{q_0} \otimes I_{q_1}  \otimes H     \mathop{\otimes} I_{q_r} )
  \end{eqnarray*}
for the case when $i=2$, and the other cases of $i=0,1,3$ can be defined likewise;
  and \isa{mat_add A B} and \isa{mat_mult A B} stand for
  the addition and multiplication of matrices resp.
                           The parameters $H$ and $\mathrm{Delta} $ (i.e. $\sigma_x$ in the source code), etc. are  of type \isa{Mat}, whose concrete values are
                            given in a text file named ``param.txt''. By applying \isa{quantum_oracle}, the order relation is decided, and after the result is returned to Isabelle/HOL, the proof is completed.
But the loop invariant of the while-loop body must be provided manually during the verification, because there has not been
 any work on synthesizing invariants for quantum programs available at this moment. 

\subsection{Quantum Phase Estimation}
The other example is the so-called quantum phase estimation, the core of Shor's algorithm as described in Algorithm~\ref{fig:QPE}. The quantum phase estimation algorithm also plays a key role in other interesting problems such as the order-finding problem. The aim of the quantum phase estimation algorithm is to find the approximation of the eigenvalues of a unitary operator under certain circumstances. Suppose a unitary operator U has an eigenvector $\ket u$  with eigenvalue $e^{2\pi i \varphi}$, where the value of $\varphi$ is unknown. The goal is to estimate $\varphi$. The quantum phase estimation procedure is not a complete quantum algorithm itself but a kind of  'subroutine' or 'module'. So to perform the estimation we assume that we have available black boxes  capable of preparing the state $\ket u$ and performing  the controlled-$U^{2^j}$, for suitable non-negative integers $j$.

The procedure uses two registers. The first register contains $n$ qubits initially in the state $\ket 0$. The choosing of $n$ depends on the number of digits of accuracy we wish to have in our estimation for $\varphi$, and the probability we wish with  the quantum phase estimation procedure to be successful.
Also, for simplicity, we choose $n=2$. The second register  $ q$ in the program   begins in the state $\ket u$, and contains as many qubits as necessary to store $\ket u$.

As presented in Algorithm~\ref{fig:QPE},
Phase estimation is performed in two stages. First, we perform quantum Fourier transformation on the first register and controlled-$U$ operations on the second register, with $U$ raised to successive powers of two. The final state of the first register after a tedious calculation becomes:
 \begin{align}
\frac{1}{2^{n/2}} \left( \ket 0  + e^{2\pi i 2^{n-1} \varphi } \ket 1 \right) \left( \ket 0 + e^{2\pi i 2^{n-2} \varphi} \ket 1 \right) \cdots \left( \ket 0 + e^{2\pi i 2^0 \varphi} \ket 1 \right )  \label{state0}
\end{align}
Supposing $\varphi$ is expressed exactly in $n$ bits, as $\varphi =0.\varphi_1 \cdots \varphi_n$, then the state {\ref{state0}} may be rewritten
{ \[\begin{array}{l}
 \frac{1}{2^{n/2}} \left( \ket 0  + e^{2\pi i 0. \varphi_n } \ket 1 \right) \left( \ket 0 + e^{2\pi i 0. \varphi_{n-1} \varphi_t} \ket 1 \right) \cdots \left( \ket 0 + e^{2\pi i 0. \varphi_1 \varphi_2 \cdots \varphi_n} \ket 1 \right )
\end{array} \] }\\
So the second stage is to apply the inverse quantum Fourier transform on the first register and then to read out the state of the first register by doing a measurement in the computational basis which can gives us $\varphi$ exactly.

It is easy to see that the precondition  can be written as :
{ \[\begin{array}{l}
P=\mathop{\bigotimes}\limits_{i=0}^{n-1} I_{q_i} \otimes \ket u \bra u\\
\end{array} \] }\\
and the postcondition:
{ \[\begin{array}{l}
Q=\ket \varphi \bra \varphi\otimes \ket u \bra u
\end{array} \] }\\
and  we need to prove   \isa{valid\ P\ S\ Q}, which can be transformed to  $P \sqsubseteq P'  $ by applying \isa{vcg}, where $S$ stands for the phase estimation algorithm and $P'$ for the deduced precondition of $S$ with respect to $Q$. The order relation $P \sqsubseteq P'  $ can then be decided by applying \isa{quantum_oracle}. Finally, we prove that \isa{valid\ P\ S\ Q} holds for the  phase estimation algorithm.

\begin{algorithm}[H]
\centering
{\small \[\begin{array}{l}
 q_0:=0;
 q_1:=0;
...;
 q_{n-1}:=0;\\
[2mm] q_0:=Hq_0;
q_0:=R_2q_0;
q_0:=R_3q_0;
...;
q_0:=R_nq_0;
q_0,q=U^{2^0}q_0,q;\\
q_1:=Hq_1;
q_1:=R_2q_1; q_1:=R_3q_1;
...;
q_1:=R_{n-1}q_1;
q_1,q=U^{2^1}q_1,q;\\
[2mm]
\quad ... \\
[2mm]
q_{n-1}:=Hq_{n-1};
q_{n-1},q=U^{2^{n-1}}q_{n-1},q;
q_{n-1},q=(U^{2^{n-1}})^{\dagger}q_{n-1},q;\\
q_{n-1}:=H^{\dagger}q_{n-1};
q_{n-2},q=(U^{2^{n-2}})^{\dagger}q_{n-2},q;
q_{n-2}:=R_2^{\dagger}q_{n-2};
q_{n-2}:=H^{\dagger}q_{n-2};
... \\
q_0,q=(U^{2^0})^{\dagger}q_0,q;
q_0:={R_{n-1}}^{\dagger}q_0;
q_0:={R_{n-2}}^{\dagger}q_0;
 ... \\
q_0:={R_2}^{\dagger}q_0;
q_0:=H^{\dagger}q_0;\\
\textbf{measure}\ M[q_0,q_1,...,q_{n-1}]:\bar{S}
\end{array}
\] }
\caption{Quantum Phase Estimation.}
\label{fig:QPE}
\end{algorithm}

\oomit{\section{Details on the theorem prover}
The theorem prover is developed based on Isabelle/HOl, a proof assistant for higher-order logic. And the matrix operation is done through Python and Numpy. The tool is available both for Linux and Windows and its source code file can be download from :
\begin{center}
\texttt{https://github.com/liutao2015/propitious-barnacle}
\end{center}

\section{Future Work} }


\section{Concluding remarks and future work}
In this paper,  we implemented a first theorem prover for quantum programs based on Isabelle/HOL.
The logical foundation of the theorem prover is based on
Quantum Hoare Logic introduced by one of the authors in \cite{Ying12}. In order to deal with extensive
calculations of matrices in the verification of quantum programs, we implemented related functions over matrices by calling
 the packages \emph{Numpy} and \emph{Sympy} in Python. Thus, the calculations of matrices
 during the verification can be done by calling these functions as an oracle of Isabelle/HOL.
 To demonstrate its power,
the correctness of two well-known quantum algorithms, i.e., Grover Quantum Search and Quantum Phase Estimation are
proved using the theorem prover, which are the first mechanized proofs for both of them.

The tool and the case studies are available both for Linux and Windows at
\begin{center}
\texttt{https://github.com/liutao2015/propitious-barnacle}.
\end{center}

Same as in the verification of classical programs using \emph{Floy-Hoare-Naur} inductive assertion method, the hardest part is still
\emph{invariant generation} and \emph{termination analysis} (for total correctnesss). Up to now, little work on invariant generation of
 quantum programs is available, therefore we have to provide an invariant manually during the verification using the prover. So, it is certainly of significance
 to investigate the problem of invariant generation of quantum programs. Besides, it deserves to consider
 the total correctness of quantum programs, and therefore the termination problem, although some work has been done \cite{LYY14,LY14}.
 In addition, it is expected  to improve the automation of the QHL prover to ease the user.

 \oomit{Since we oop invariant which is still a kind of matrix need to be given manually in our prover.In other words, the prover could not generate the loop invariant automaticlly.And that is what we will work for in the future.

Duing the fact that what we have done is the first theorem prover for quantum hoare logic in the world, and it is  the first version, we encounterd some problems that would be solved later and there are still a few aspects can be improved.The proving process of quantum program is some different from that of classical program.The prover of calssical program tends to do much about logic inference, while the prover of quantum program needs much calculation in which the matrix operations are involed, especially the complex matrix operations . However, several theorem provers existed now including Isabelle/HOL and Coq support little about matrix  claculation. And it is still hard to input the matrix in Isabelle/HOL and Coq. So we  declare a new type \isa{Mat} to represent matrix and defines its meaning by a set of axioms. When proving the Grover Search algorthm and Shor's algorthm, we just take advantage of the power of automatic inference of Isabelle/HOL to  record the computing process and perform the concret calculation with outside module we developed by Python as well as Numpy and Sympy. So we will try to develop a new tool or an extension to Isabelle/HOL or Coq that suits the development of the prover in the future. }

\bibliographystyle{plain}
\bibliography{bibForProver}

\end{document}